# Detection Implicit Links and G-betweenness

M.I. Zhenirovskyy[1], D.V. Lande[2], A.A. Snarskii[2]

[1]N. N. Bogolyubov Institute for Theoretical Physics, Kiev, Ukraine

[2]National Technical University of Ukraine (KPI), Kiev, Ukraine

A concept of implicit links for Complex Networks has been defined and a new value – cohesion factor, which allows to evaluate numerically the presence of such connection between any two nodes, has been introduced. We introduce a generalization of such characteristics as the betweenness, which allows to rank network nodes in more details. The effectiveness of the proposed concepts is shown by the numerical examples.

PACS number: 89.75.Hc, 89.65.-s

## 1. Introduction

Complex networks theory that has been actively developing recently [1, 2], allows us to understand and numerically characterize many properties of the surrounding world, such as social and economic links, traffic of transport, energy, information and relations in biological systems, and many other things.

One of the major problems confronting the world community since September 2001 is the prevention of terrorist attacks. It requires to solve a number of specific tasks, one of which - the identification of linkages between members of terrorist groups, which as a rule are being carefully hidden. Methods of the theory of complex networks allow to set the task correctly and to solve it in one or another approach [3, 4].

In our work we define the concept of so-called implicit (intentionally hidden) relationships and introduce a new value – cohesion factor, which allows to evaluate numerically the presence of such a connection between any two network members. In addition we introduce a generalization of such node characteristic as the site betweenneess [5] - G-betweenneess, which allows, in some cases, to rank network nodes in more details. The effectiveness of the proposed concepts is shown by the numerical examples.

---

[1]mzhenirovskyy@gmail.com

In the Appendix at the end of the work a method for calculating the conductivity between network nodes is shown.

## 2. Cohesion factor

Lets look at complex network which nodes are the objects of the real world (people, banks, servers, etc.), and the weight connection between them - characteristics of the interaction. For example, the number of contacts per unit time, the joint capital, etc. Figure 1 shows two versions of connections between nodes $i$ and $j$.

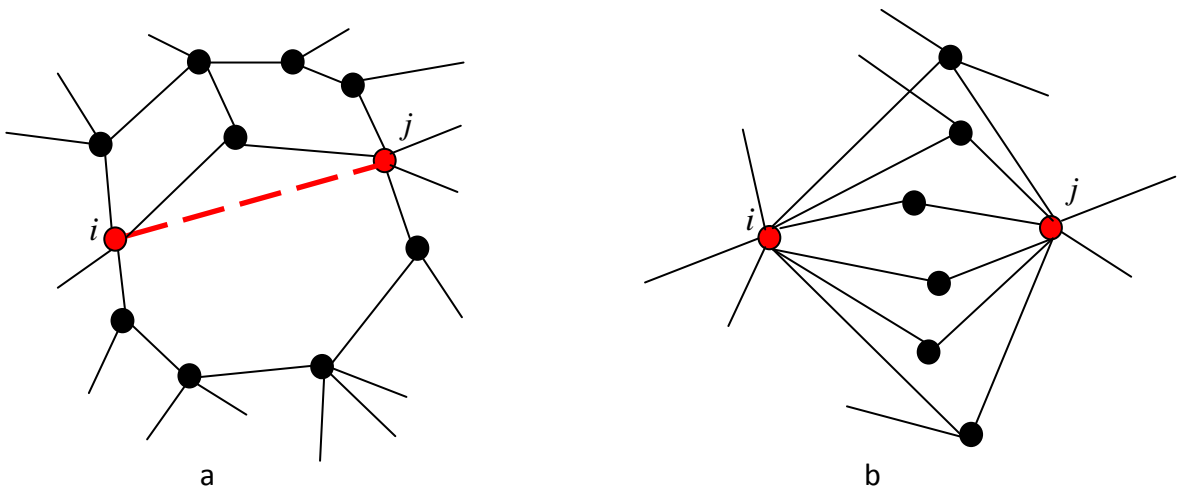

a　　　　　　　　　　　　　　　　　b

Fig. 1. Part of a network containing nodes $i$ and $j$. a - there is a clear link between $i$ the $j$ node (red dotted line), b - nodes $i$ and $j$ are directly unrelated, but between them there are many links through one neighbour.

In the first case (Fig. 1) nodes $i$ and $j$ are directly linked, in the adjacency matrix $\mathbf{A}$ of such network element $A_{ij}$ is equal to the weight of the connection. In the second case (Fig. 1b) there is no direct link between $i$ nodes $j$, but there are many links with length two. In the adjacency matrix, a network element $A_{ij} = 0$, however, it often does not mean that the nodes $i$ and $j$ are linked "weaker" in the second case than in the first one. Thus, for example, if we are talking about the traffic, then, despite the existence of a direct connection in the first scenario, the total traffic between nodes $i$ and $j$ in the second option may be substantially greater.

To determine the "strength" between nodes, which would take into account not only direct links. It is necessary to formulate a criterion which takes into account not only the existence of direct links and relationships through the nearest neighbours, but also their number and topology. Such a well-known characteristic as the shortest path, is inconvenient in this case, since the existence of a set of long paths can give a greater contribution to the "strength" of communication, rather than the shortest one [6].

Let's set the conductance (the inverse value of electrical resistance) of each connection as a quantity numerically equal to the weight of the connection. We characterize the force of connection between nodes $i$ and $j$ cohesion factor - $K_{ij}$. The cohesion factor is defined as normalized average of the net value of the conductance between nodes $i$ and $j$. During calculation of the average value only directly connected nodes shall be taken into consideration.. The weight of each bond equals its conductance, and calculations are carried out according to well-known rules of Kirchhoff [6, 7] (see the Appendix).

With this definition of "power" all links shall be taken into consideration, not just the direct or shortest routes, their topology and weight. And if for some pair of nodes that are not directly related, cohesion factor is more than one, then these nodes will be called implicitly linked. Thus, for example, if all weights of networks represented in Fig. 1 are the same, then according to the definition of cohesion, the strength of links between nodes $i$ and $j$ in the second version is more, as it appears from the general considerations.

Let's consider a network characterized by its adjacency matrix, so that the $A_{ij}$ is weight of link - conductance between nodes $i$ and $j$. Let's introduce the matrix $\mathbf{B}$ defined as follows:

$$B_{ij} = \begin{cases} 1, & A_{ij} \neq 0 \\ 0 & A_{ij} = 0 \end{cases} \tag{1}$$

Average conductance $\langle G \rangle$ by explicit links is

$$\langle G \rangle = \frac{1}{\sum_{i,j} B_{ij}} \sum_{i,j}^{N} B_{ij} \cdot G_{ij}, \tag{2}$$

Where $G_{i,j}$ is conductance between nodes $i$ and $j$ which takes into account all the connections. Cohesion factor $K_{i,j}$ for the nodes $i$ and $j$ is defined as follows:

$$K_{ij} = \frac{G_{ij}}{\langle G \rangle}. \qquad (3)$$

To illustrate introduced concepts we consider a model example - Figure 2. Let's write the incidence matrix of this network which is shown in Figure 2:

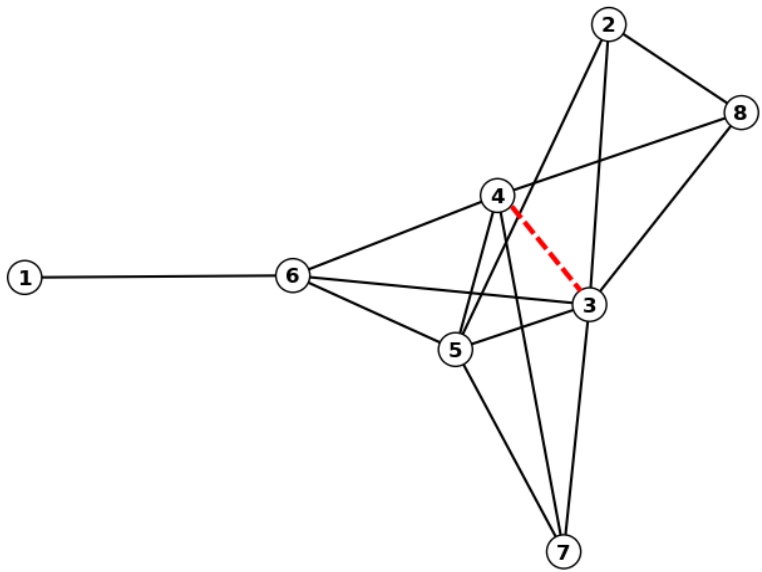

Fig. 2. Model of the network with 8 nodes. The weights of all links (conductivity) are unity. Red dotted line shows the implicit connection.

$$A = \begin{pmatrix} 0 & 0 & 0 & 0 & 0 & 1 & 0 & 0 \\ 0 & 0 & 1 & 0 & 1 & 0 & 0 & 1 \\ 0 & 1 & 0 & 0 & 1 & 1 & 1 & 1 \\ 0 & 0 & 0 & 0 & 1 & 1 & 1 & 1 \\ 0 & 1 & 1 & 1 & 0 & 1 & 1 & 0 \\ 1 & 0 & 1 & 1 & 1 & 0 & 0 & 0 \\ 0 & 0 & 1 & 1 & 1 & 0 & 0 & 0 \\ 0 & 1 & 1 & 1 & 0 & 0 & 0 & 0 \end{pmatrix}. \qquad (4)$$

In this example $\mathbf{A} = \mathbf{B}$. Calculating the conductivity matrix $\mathbf{G}$ (see the Appendix) we find:

$$G = \begin{pmatrix} 0 & 0.58 & 0.68 & 0.67 & 0.69 & 1.00 & 0.60 & 0.58 \\ 0.58 & 0 & 2.22 & 1.56 & 2.06 & 1.38 & 1.38 & 1.91 \\ 0.68 & 2.22 & 0 & 2.15 & 2.87 & 2.13 & 2.13 & 2.15 \\ 0.67 & 1.56 & 2.15 & 0 & 2.55 & 2.06 & 2.06 & 1.89 \\ 0.69 & 2.06 & 2.87 & 2.55 & 0 & 2.24 & 2.24 & 1.76 \\ 1.00 & 1.38 & 2.13 & 2.06 & 2.24 & 0 & 1.50 & 1.39 \\ 0.60 & 1.38 & 2.13 & 2.06 & 2.24 & 1.50 & 0 & 1.39 \\ 0.58 & 1.91 & 2.15 & 1.89 & 1.76 & 1.39 & 1.39 & 0 \end{pmatrix}, \qquad (5)$$

From (2) for the average conductance we directly obtain $\langle G \rangle = 2.11$. And if we now draw attention to the matrix element $G_{34} = 2.15$, we can see that on the one hand $A_{34} = 0$, so between nodes 4 and 3, there is no direct connection (Fig. 2). But the conductivity between these nodes is greater than the average one for obvious connections. I.e. cohesion factor (3) for nodes 3 and 4 $K_{34} = G_{34}/\langle G \rangle = 1.02$ is more than one, and it's logical to assume that the connection between these nodes is implicit.

## 3. G-betweenness.

Let's consider another example of the network (see Fig. 3).

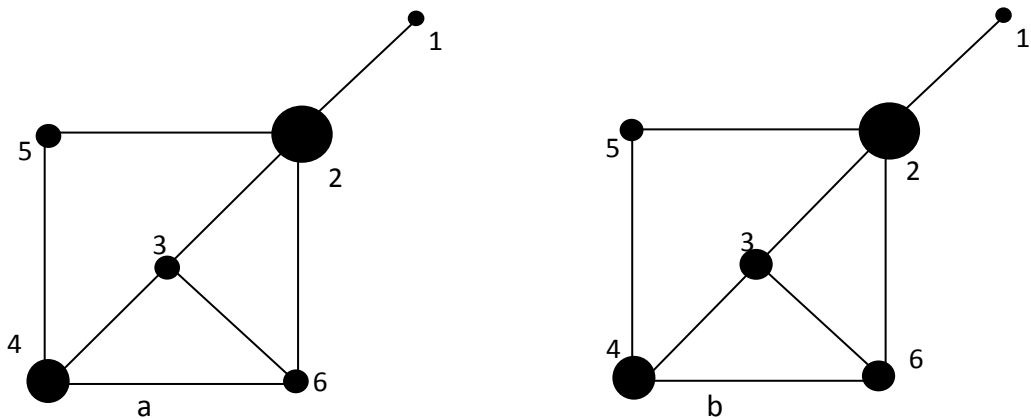

Fig. 3. The simplest network, which clearly shows a - nodes 3 and 6 are identical and fundamentally differ from node 5, while all these units - 3, 6 and 5 have the same value of betweenness, b-entered (introduced) new characteristic - G-betweenness gives the same values for nodes 3 and 6, and different one for node 5.

Let's calculate betweenness for each node of the network. Betweenness describes the role of the node as an intermediary in establishing links between other nodes, its numerical value is in proportion to the number of shortest paths for different nodes, passing through this node [5]. For a node $n$ intervention $B(n)$ is defined as:

$$B(n) = \sum_{i \neq j} \frac{B(i,n,j)}{B(i,j)}, \qquad (6)$$

where $B(i,j)$ - a number of shortest paths between nodes $i$ and $j$, $B(i,n,j)$ that portion of them, which passes through the node $n$. For our example we get: $B_1 = 0$, $B_2 = 10$, $B_3 = 4/3$, $B_4 = 2$, $B_5 = 4/3$, $B_6 = 4/3$. We have got the same values for nodes 3, 5 and 6. Nodes 3 and 6 are really equivalent to each other, but node 5 differs from them. Let's consider the matrix SP - a short path for our network (7). For instance, short path from node 3 to any other node is the same as from node 6. But this is not the case for 5 node.

$$\mathbf{SP} = \begin{pmatrix} 0 & 1 & 2 & 3 & 2 & 2 \\ 1 & 0 & 1 & 2 & 1 & 1 \\ 2 & 1 & 0 & 1 & 2 & 1 \\ 3 & 2 & 1 & 0 & 1 & 1 \\ 2 & 1 & 2 & 1 & 0 & 2 \\ 2 & 1 & 1 & 1 & 2 & 0 \end{pmatrix}. \qquad (7)$$

I.e. such a standard definition of betweenness - (6) does not detect all differences between the nodes.

We introduce a new characterisitic of a node - G-betweenness, which also characterizes the node as a mediator, but in the same time takes into consideration all, not just the shortest paths between different nodes, passing through this node. G-betweenness determines how the total conductivity between the nodes has changed when you remove the node from the network.

We calculate, for example G-betweenness for the network node 5, shown in Figure 3. First of all, we find a conductivity matrix between all network nodes (see Ex.4):

$$G = \begin{pmatrix} 0 & 1.00 & 0.65 & 0.60 & 0.60 & 0.65 \\ 1.00 & 0 & 1.85 & 1.50 & 1.50 & 1.85 \\ 0.65 & 1.85 & 0 & 1.85 & 1.14 & 2.00 \\ 0.60 & 1.50 & 1.85 & 0 & 1.50 & 1.85 \\ 0.60 & 1.50 & 1.14 & 1.50 & 0 & 1.14 \\ 0.65 & 1.85 & 2.00 & 1.85 & 1.14 & 0 \end{pmatrix}. \qquad (8)$$

Now we will remove the fifth node from the network – i.e. assign zero weight (zero conductivity) to all the bonds of the fifth node . And we again find the matrix conductivity between all nodes.

$$G5 = \begin{pmatrix} 0 & 1.00 & 0.62 & 0.50 & 0 & 0.62 \\ 1.00 & 0 & 1.60 & 1.00 & 0 & 1.60 \\ 0.62 & 1.60 & 0 & 1.60 & 0 & 2.00 \\ 0.50 & 1.00 & 1.60 & 0 & 0 & 1.60 \\ 0 & 0 & 0 & 0 & 0 & 0 \\ 0.62 & 1.60 & 2.00 & 1.60 & 0 & 0 \end{pmatrix}. \qquad (9)$$

Let's define the matrix of the relative difference **GG** as follows:

$$GG_{ij} = \frac{G_{ij} - G1_{ij}}{G_{ij}}. \qquad (10)$$

In our case:

$$\mathbf{GG} = \begin{pmatrix} 0 & 0 & 0.05 & 0.17 & 1.00 & 0.05 \\ 0 & 0 & 0.13 & 0.33 & 1.00 & 0.13 \\ 0.05 & 0.13 & 0 & 0.13 & 1.00 & 0 \\ 0.17 & 0.33 & 0.13 & 0 & 1.00 & 0.13 \\ 1.00 & 1.00 & 1.00 & 1.00 & 0 & 1.00 \\ 0.05 & 0.13 & 0 & 0.13 & 1.00 & 0 \end{pmatrix}. \tag{11}$$

And finally for the fifth node, we obtain:

$$GB_5 = \sum_{i \neq 5} \sum_{j \neq 5} GG_{ij} = 2.27. \tag{12}$$

With similar calculation for G-betweenness for all nodes in our network, we obtain: $GB_1 = 0$, $GB_2 = 11.82$, $GB_3 = 3.14$, $GB_4 = 4.45$, $GB_5 = 2.27$, $GB_6 = 3.14$. Now, we can clearly see that nodes 3 and 6 are equivalent, and node 5 is different to them.

The expression for the determination of G-betweenness k node can be written as follows:

$$GB_k = \sum_{i \neq k} \sum_{j \neq k} \frac{G_{ij} - Gk_{ij}}{G_{ij}}, \tag{13}$$

where $G$ – conduction matrix of the network, $Gk$ - conduction matrix of the network with removed k node.

## Conclusion

In this paper we introduce two network characteristics - cohesion factor and G-betweenness. The first of them reveals the implicit connection between network elements. In particular, to determine the nodes that are not directly related (perhaps this is a direct link deliberately hidden), but have a lot of neighbourhood connections. The second concept of G-betweenness - allows to find those nodes, which act as intermediaries in establishing links between other nodes. In this case, in contrast to, the standard characteristic Betweenness, the introduced characteristic G-betweenness (similar to the coefficient factor) takes into account not only single path (the length of one link), but also longer ones, through the neighbours. Both introduced characteristics use the analogy between weight of (connection, links, relation) and electrical conductivity.

**Appendix**

To determine the conductance between any two network nodes, we use a method of nodal potentials that is well-known in the theory of electrical circuits (see for example [7]). This method provides solution for a system of linear algebraic equations (SLAE) for the Laplace matrix and driving current vector w. At the same time the vector of nodal potentials acts as the required vector. It should be noted that as usually in physics, one arbitrary node is assigned a zero potential, after that the column and row number of the zero-node are just removed from the matrix of Laplace, that is why the matrix of Laplace, in this case has dimension one less than the incidence matrix. Elements of the matrix Laplace are defined as follows:

$$L_{ij} = \begin{cases} -A_{ij}, & i \neq j \\ \sum_{j \neq i} A_{ij}, & i = j \end{cases}, \qquad (\text{Ex.1})$$

where $i, j = 1, 2, \ldots, N-1$ ($N$ - size of the incidence matrix). Note that the network with undirected links are considered, i.e. $A_{ij} = A_{ji}$. After that it's necessary to write the driving current vector $\mathbf{I}$. For example, we are looking for the resistance between nodes $i$ and $j$. Then the elements of the vector $\mathbf{I}$ are shown as follows:

$$I_k = \begin{cases} 0, & k \neq i, k \neq j \\ 1, & k = i \\ -1, & k = j \end{cases}. \qquad (\text{Ex.2})$$

This corresponds to the unit current flowing into the node $i$ and the resulting current from the node $j$. In matrix form our SLAE is as follows:

$$\mathbf{L} \cdot \mathbf{V} = \mathbf{I}, \quad (Ex.3)$$

where $\mathbf{V}$ - the required vector of nodal potentials. And consequently the conductance between nodes $i$ and $j$ is defined as $G_{ij} = 1/|V_i - V_j|$.

In the numerical solution of (3) and large $N$ you have to use one of the methods (see e.g. [8]), which takes into account the symmetry and sparseness that in most practical cases is true for Laplace matrix. If the system is not very big and well-conditioned, then we can immediately write the solution for the conductance between nodes $i$ and $j$.

$$G_{ij} = \frac{\det \mathbf{L}}{\det L_{(i+j)(i+j)}} \quad (Ex.4)$$

Here $L_{(i+j)(i+j)}$ - a matrix $\mathbf{L}$ minor, which is calculated as follows: line $i$ is added to the line $k$ and then deleted, the column $i$ is added to the column $j$, and then also deleted. If one of the indexes is zero, then simply delete the column and row corresponding to the nonzero indexes.